\documentclass[twocolumn]{aastex63}

\usepackage[fleqn]{amsmath}
\renewcommand{\ng}{NGC\ 4151 }
\newcommand{\ngs}{NGC\ 4151}
\newcommand{\h}{$H_0$ }

\newcommand{\hst}{{\it HST} }
\newcommand{\hsts}{{\it HST}}
\newcommand{\hstw}{{\it F350LP} }
\newcommand{\hstv}{{\it F555W} }
\newcommand{\hsti}{{\it F814W} }
\newcommand{\hsth}{{\it F160W} }
\newcommand{\hstws}{{\it F350LP}}
\newcommand{\hstvs}{{\it F555W}}
\newcommand{\hstis}{{\it F814W}}
\newcommand{\hsths}{{\it F160W}}

\newcommand{\B}{{\it B}-band}                                                             
\newcommand{\V}{{\it V}-band}                                                             
\newcommand{\R}{{\it R}-band}                                                             
\newcommand{\I}{{\it I}-band} 
   
\newcommand{\ignore}[1]{}

\newcommand{\msun}{\mbox{$M_\odot$}}

\usepackage{xcolor}

\makeatletter
\newcommand*{\rom}[1]{\expandafter\@slowromancap\romannumeral #1@}
\makeatother

\shorttitle{Distance to the Seyfert Galaxy \ngs}
\shortauthors{Yuan et al.}

\begin{document}
\title{The Cepheid Distance to the Seyfert 1 Galaxy NGC\,4151}

\author[0000-0001-9420-6525]{W.~Yuan}
\affiliation{Department of Physics \& Astronomy, Johns Hopkins University, Baltimore, MD 21218, USA}

\author[0000-0002-9113-7162]{M.~M.~Fausnaugh}
\affiliation{Department of Astronomy, The Ohio State University, 140 W 18th Ave, Columbus, OH 43210, USA}
\affiliation{\ignore{}Kavli Institute for Space and Astrophysics Research,  Massachusetts Institute of Technology,\\ 
77 Massachusetts Avenue, Cambridge, MA 02139, USA}

\author[0000-0002-4312-7015]{S.~L.~Hoffmann}
\affiliation{Space Telescope Science Institute, 3700 San Martin Drive, Baltimore, MD 21218, USA}

\author[0000-0002-1775-4859]{L.~M.~Macri}
\affiliation{George P.\ and Cynthia W.\ Mitchell Institute for Fundamental Physics and Astronomy, Department of Physics and Astronomy,\\Texas A\&M University, College Station, TX 77843, USA}

\author[0000-0001-6481-5397]{B.~M.~Peterson}
\affiliation{Department of Astronomy, The Ohio State University, 140 W 18th Ave, Columbus, OH 43210, USA}
\affiliation{Center for Cosmology and AstroParticle Physics, The Ohio State University, 191 West Woodruff Ave, Columbus, OH 43210, USA}
\affiliation{Space Telescope Science Institute, 3700 San Martin Drive, Baltimore, MD 21218, USA}

\author[0000-0002-6124-1196]{A.~G.~Riess}
\affiliation{Department of Physics \& Astronomy, Johns Hopkins University, Baltimore, MD 21218, USA}
\affiliation{Space Telescope Science Institute, 3700 San Martin Drive, Baltimore, MD 21218, USA}

\author[0000-0002-2816-5398]{M.~C.~Bentz}
\affiliation{\ignore{Georgia}Department of Physics and Astronomy, Georgia State University, 25 Park Place, Suite 605, Atlanta, GA 30303, USA}

\author{J.~S.~Brown}
\affiliation{\ignore{UCSC}UCO/Lick Observatory, University of California, Santa Cruz, CA 95064, USA}

\author[0000-0001-9931-8681]{E.~Dalla~Bont\`{a}}
\affiliation{\ignore{Padova}Dipartimento di Fisica e Astronomia ``G. Galilei,'' Universit\`{a} di Padova, Vicolo dell'Osservatorio 3, I-35122 Padova, Italy}
\affiliation{\ignore{INAF}INAF-Osservatorio Astronomico di Padova, Vicolo dell'Osservatorio 5 I-35122, Padova, Italy}

\author[0000-0003-4949-7217]{R.~I.~Davies}
\affiliation{\ignore{MPE}Max Planck Institut f\"ur extraterrestrische Physik, Postfach 1312, D-85741, Garching, Germany}

\author[0000-0003-3242-7052]{G.~De~Rosa}
\affiliation{Space Telescope Science Institute, 3700 San Martin Drive, Baltimore, MD 21218, USA}

\author[0000-0002-8224-1128]{L.~Ferrarese}
\affiliation{\ignore{Victoria}NRC Herzberg Astronomy and Astrophysics, National Research Council, 5071 West Saanich Road, Victoria, BC V9E 2E7, Canada}

\author[0000-0001-9920-6057]{C.~J.~Grier}
\affiliation{Department of Astronomy, The Ohio State University, 140 W 18th Ave, Columbus, OH 43210, USA}
\affiliation{\ignore{Steward}Steward Observatory, University of Arizona, 933 North Cherry Avenue, Tucson, AZ 85721, USA}

\author[0000-0002-4457-5733]{E.~K.~S.~Hicks}
\affiliation{\ignore{Anchorage}Department of Physics and Astronomy, University of Alaska Anchorage, AK 99508, USA}

\author[0000-0003-0017-349X]{C.~A.~Onken}
\affiliation{\ignore{ANU}Research School of Astronomy and Astrophysics, Australian National University, Canberra, ACT 2611, Australia}
\affiliation{Australian Research Council (ARC) Centre of Excellence in All-sky Astrophysics (CAASTRO)}

\author[0000-0003-1435-3053]{R.~W.~Pogge}
\affiliation{Department of Astronomy, The Ohio State University, 140 W 18th Ave, Columbus, OH 43210, USA}
\affiliation{Center for Cosmology and AstroParticle Physics, The Ohio State University, 191 West Woodruff Ave, Columbus, OH 43210, USA}

\author[0000-0003-1772-0023]{T.~Storchi-Bergmann}
\affiliation{\ignore{UFRGS}Departamento de Astronomia, Universidade Federal do Rio Grande do Sul, Av. Bento Goncalves 9500, 91501 Porto Alegre, RS, Brazil}

\author[0000-0001-9191-9837]{M.~Vestergaard}
\affiliation{\ignore{Dark}DARK Niels Bohr Institute, University of Copenhagen, Jagtvej 128, 2200 Copenhagen N, Denmark}
\affiliation{\ignore{Steward}Steward Observatory, University of Arizona, 933 North Cherry Avenue, Tucson, AZ 85721, USA}
  
\begin{abstract}
We derive a distance of $15.8\pm0.4$~Mpc to the archetypical Seyfert 1 galaxy \ng based on the near-infrared Cepheid Period--Luminosity relation and new {\it Hubble Space Telescope} multiband imaging.  This distance determination, based on measurements of 35 long-period ($P\!>\!25$d) Cepheids, will support the absolute calibration of the supermassive black hole mass in this system, as well as studies of the dynamics of the feedback or feeding of its active galactic nucleus.\\
\end{abstract}

\ \par

\section{Introduction}

\ \par

Local active galactic nuclei (AGNs) provide unique laboratories to study the co-evolution of galaxies and their supermassive black holes (SMBHs).  At redshifts $z\!<\!0.005$ ($D\!\lesssim\!20$ Mpc), the host galaxies are well resolved, enabling detailed studies of the structure and dynamics of the their interstellar gas and any connection to the AGN through feedback or feeding. Additionally, since the sphere of influence of the central supermassive black holes is spatially resolved in these local systems, the black hole mass can be measured using resolved dynamical methods. Local AGNs, therefore, are not only central to our understanding of SMBH scaling relations in AGNs, but are also crucial in calibrating SMBH mass measurements based on methods (e.g., reverberation mapping) applicable at higher redshifts (see \citealt{Ferrarese2005,McConnell2013,Sahu2019,Zubovas2019} for reviews).

However, these local AGNs are subject to a single systematic uncertainty that makes all physical parameters highly uncertain: their distances. With accurate knowledge of the local Hubble constant $H_0$ (currently better than a few
percent, though tension with results from the cosmic microwave background  and baryon acoustic oscillations remains; \citealt{Riess2019}), the uncertainty of extragalactic distances is principally limited by peculiar velocities relative to the Hubble flow.   At very low redshifts, the peculiar velocities can be large compared to the cosmological expansion rate.  Even for galaxies as far away as 20 Mpc, peculiar velocities can introduce errors of 20\% or more in their redshift-based distances and, of course, the problem is more severe for nearer galaxies.

Redshift-independent distances for the late-type hosts of local AGNs are available only  through the Tully--Fisher relation \citep{Tully1977}, but nevertheless are typically uncertain by an irreducible 20\%, and more for galaxies at lower inclination. Without an accurate distance, measurements of many other parameters are of limited utility, hampering our ability to understand the impact of the AGN on the evolution of the host galaxy.  Thus, we find ourselves in a paradoxical situation for these local AGNs:  even while they are the best systems for detailed studies of the structure, dynamics, and energetics (including luminosity measures) of the AGN and host galaxy, these are precisely the objects for which a lack of accurate distances makes all physical parameters highly uncertain. 

Understanding the masses of SMBH in nearby AGNs also depends on accurate distance measurements.  AGN black hole masses are generally best determined by reverberation mapping \citep[RM;][]{Peterson2014}. As it is usually applied, RM yields black hole mass uncertainties at $\sim0.3$--$0.4$ dex, but careful modeling of high-quality data can reduce the uncertainties to $0.1$--$0.2$ dex \citep{Pancoast2014b,Grier2017} and are distance-independent.  However, verification of RM-based masses requires independent measurements using other techniques such as modeling stellar or gas dynamics and, more recently, interferometry \citep{Sturm2018}, that are distance dependent and, for the most part, applicable only to the nearest AGNs. Accurate distances to nearby AGNs are thus crucial for verification of RM-based masses by other techniques, particularly as the mass measurements themselves become increasingly precise.

For galaxies out to $\sim$50\,Mpc, Cepheid variables provide an excellent means of distance determination, enabled by the Period--Luminosity Relation (PLR) or Leavitt Law \citep{Leavitt1912}. Recently, \citet{Bentz2019} applied this method at optical wavelengths to the  Seyfert 1 galaxy NGC$\,$6814, in order to compare the SMBH mass inferred from dynamical modeling and reverberation mapping.  In this contribution, we employ Cepheid variables to determine the distance to the well-known nearby AGN \ngs.

\ng is  one of the best-studied AGNs at all wavelengths \citep[e.g.,][]{Ulrich2000}. Yet, despite the intensive scrutiny that it has received on account of its proximity, we do not know its distance to any acceptable degree of accuracy; indeed, the most extreme distance estimates differ by more than a factor of seven. The Extragalactic Distance Database \citep{Tully2009} quotes a distance for \ng of 11.2 Mpc based on the average distances of its fellow group members. Only four galaxies contribute to this group-averaged distance, however, with individual distances ranging from 3.9 to 34\,Mpc based on the Tully--Fisher line width--luminosity correlation. The object with the smallest estimated distance is \ng itself at 3.9 Mpc; however, the total galaxy luminosity has not been corrected for the large contribution from the AGN itself, which would cause the galaxy to appear brighter, and therefore nearer, than it actually is. The situation is further confused by the various adopted values in recent studies. We summarize the literature distance measurements to \ng and adopted values in recent works in Table~\ref{tab_dis}.

\begin{deluxetable*}{cccl}
\tablecaption{Measured/Adopted Distances to \ng in Literature\label{tab_dis}}
\tablewidth{0pt}
\tablehead{
  \colhead{Reference} & \colhead{Distance} & \colhead{Uncertainty} & \colhead{Method} \\ \cline{2-3}
  & \multicolumn{2}{c}{(Mpc)} &
}
\startdata
\citet{Tully2009} &      3.9 &   0.7 & Tully--Fisher relation\\
\citet{1981ApJ...248..408D} &      4.5 &      0.8 & Tully--Fisher relation\\
\citet{1985AAS...59...43B} & 4.6 & 0.9 & Tully--Fisher relation \\
\citet{1984AAS...56..381B} &      4.8 &      0.6 & Tully--Fisher relation\\
\citet{Tully2009} &      11.2 & 1.1 & Tully--Fisher relations of 4 \ng group members\\
\citet{2007AA...465...71T} &     16.1 &      3.0 & Tully--Fisher relation\\
\citet{2019MNRAS.487.3001T} & 16.6 & 1.1 & SN II-P standard candle method\\
\citet{Cackett2007} & 19.0 & 2.0 & Inter-band continuum lags and a thermal reprocessing model \\
\citet{Hoenig2014} & 19.0 & $^{+2.4}_{-2.6}$ & IR interferometric angular size and dust reverberation \\
\citet{2019MNRAS.487.3001T} & 20.0 & 1.6 & SN II-P expanding photosphere method \\
\citet{1988cng..book.....T} &     20.3 &      3.8 & Tully--Fisher relation\\
\citet{Yoshii2014} & 29.2 & 0.4 & Dust reverberation \\
\hline
mean & 14.1 & \nodata & average of the above measurements \\
standard deviation & 8.2 & \nodata & standard deviation of the above measurements \\
\hline
\citet{Hicks2008} & 13.2 & \nodata &  adopted \\
\citet{Storchi-Bergmann2010} & 13.3 &  \nodata &  adopted \\
\citet{Iserlohe2013} & 13.3 & \nodata  &  adopted \\
\citet{Onken2014} & 13.9 & \nodata  &  adopted \\
\citet{DeRosa2018} & 13.9 & \nodata  &  adopted \\
\citet{Esquej2014} & 14.9 & \nodata  &  adopted \\
\citet{Bentz2013} & 16.6 & \nodata  &  adopted \\
\citet{Burtscher2013} & 16.9 &  \nodata &  adopted \\
\citet{Kishimoto2011} & 17.6 &  \nodata &  adopted
\enddata
\tablecomments{``Adopted'' means the distance was assumed in the corresponding study.}
\end{deluxetable*}

An uncertain distance to \ng limits our understanding of the feeding and feedback processes that we can otherwise study in great detail. For example, the geometry, kinematics, and excitation (e.g., \citealt{Winge1999, Das2005, Shimono2010, Storchi-Bergmann2009}) of the spatially extended narrow-line region have been well-characterized. AGN-driven outflows have been mapped by \cite{Storchi-Bergmann2010}, using near-infrared integral field spectroscopy, down to 8 pc from the nucleus. The UV absorption spectrum \citep[e.g.,][]{Crenshaw2007} demonstrates that there is considerable kinetic energy in outflows. The calculated mass outflow rate, as well as the power of the outflow, depend on the luminosity of the emitting gas as well as on the geometry of the outflow, both strongly dependent on the distance to the galaxy. It also appears that energetic feedback is occurring on large, resolvable scales. For example, \cite{Wang2010} study the X-ray emission that fills the ``H\,{\sc i} cavity'' in the central $\sim2$\,kpc of \ng and consider four separate hypotheses on the origin of the X-rays and the energetics of the interaction between the AGN and the host galaxy. These estimates of the energetics depend linearly on distance, but they still constitute some of the best measurements of AGN feedback in any galaxy in the observable universe.

Finally, \ng is one of the few AGNs where the mass of the supermassive black hole has been measured by more than one technique: from stellar dynamics, $M_{\rm BH} = 3.76\pm 1.15 \times 10^7\,M_\odot$ \citep{Onken2014}; from gas dynamics, $M_{\rm BH} = 3.0^{+0.75}_{-2.2} \times 10^7\,M_\odot$ \citep{Hicks2008}; and from reverberation mapping, $M_{\rm BH} = 2.14^{+0.75}_{-0.55} \times 10^7\,M_\odot$, with an additional systematic uncertainty of $\sim 0.15$\,dex \citep{DeRosa2018}. The precision of the reverberation measurement is likely to improve with detailed modeling.

In this work, we derive the distance to \ng based on one of the most accurate and widely used methods, the Cepheid PLR, using data obtained with the {\it Hubble Space Telescope} (\hsts). The rest of this article is organized as follows: in \S2, we detail our observations, data reduction, and photometry; in \S3, we describe the Cepheid search; we present our results in \S4 and discuss our findings in \S5.

\section{Observations, Data Reduction, and Photometry}

\begin{figure*}
\epsscale{1.2}
\plotone{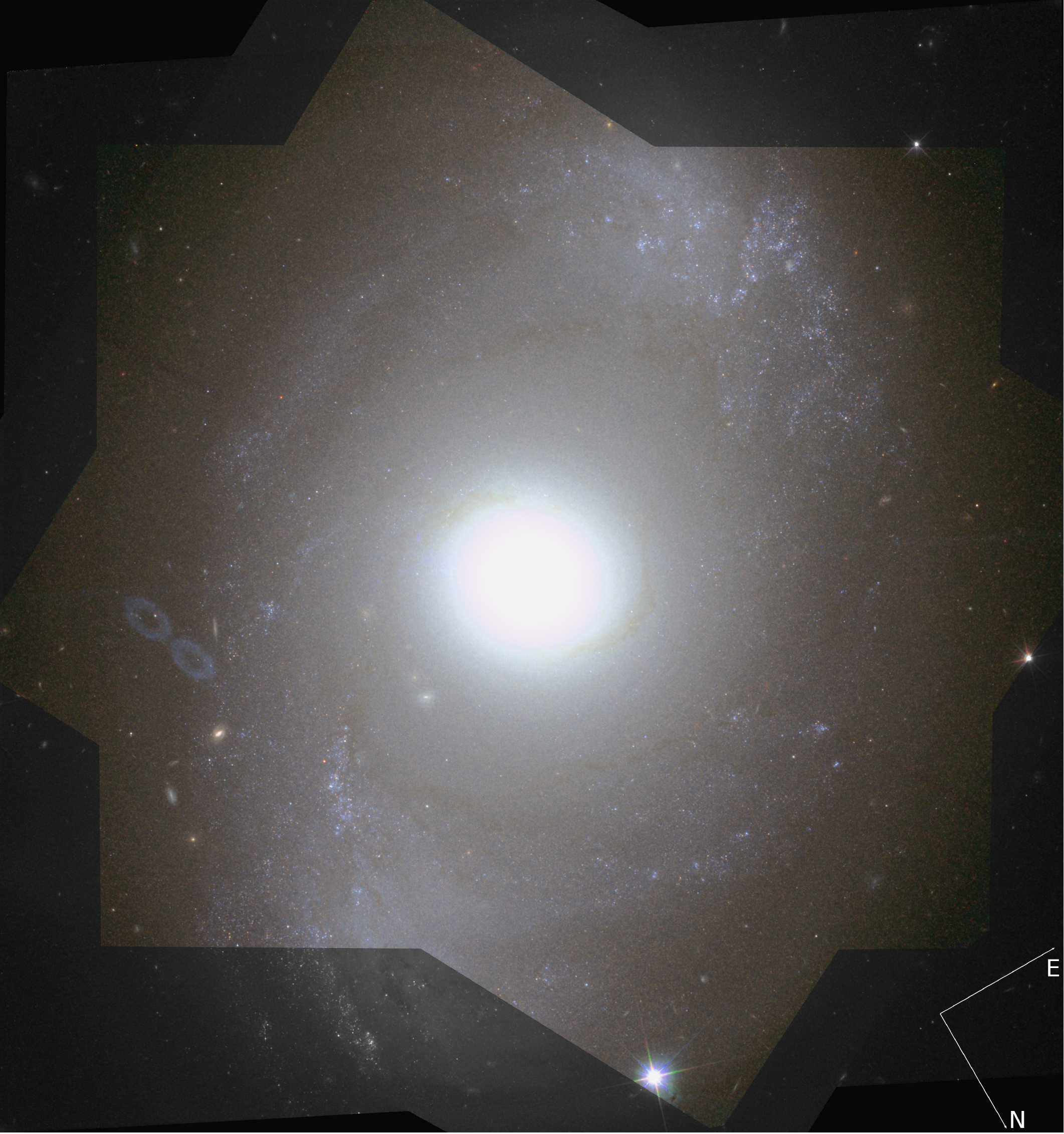}
\caption{Pseudo-color image of \ngs\ using our master frames in \hstws, \hstvs, \hstis, and \hsth as the luminance layer and the blue, green, and red channels, respectively. A square-root scaling was applied to all layers to increase visibility of the outer image. Color is only shown where all layers overlap, while grayscale is used in areas not covered in \hsths. The compass lines are $20\arcsec$ in length.\label{fig_cc}}
\end{figure*}

We acquired multiband time-series images of \ng using the Wide Field Camera 3 (WFC3) onboard \hsts. The galaxy's optically bright disk  fits well within the field of view of both the UVIS and IR detectors of this instrument. Thanks to the exceptional resolving power of \hsts, hundreds of thousands of bright stars, including long-period Cepheids, can be photometrically measured in the images. A color composite of \ng from these \hst observations is shown in Fig.~\ref{fig_cc}. We reduced these \hst data and performed photometry using {\tt DAOPHOT}~\citep{Stetson1987} and related programs. We emphasize that the analysis tools, software, and parameter settings used for this work are not the same as those used by the SH0ES project \citep{Hoffmann2016, Riess2016}. These differences pertain to the methods used for image registration, point source catalog creation, photometry, PSF modeling, and Cepheid selection. A detailed comparison study between these two approaches is beyond the scope of this work (though we compared the distance derived here with that from a private communication by the SH0ES Team and found excellent agreement).

\subsection{Observations}

In order to search for Cepheids and measure their periods, we made use of the long-pass optical filter \hstw to phase the light curves with 12 epochs across a baseline of 74 days, in the middle of which the telescope's orientation angle changed once. The \hstw filter transmission curve is significantly broader in wavelength than the combination of \hstv and \hsti filters, boosting the Cepheid discovery efficiency by a factor of $\sim2.5$ compared to the filters used traditionally \citep{Hoffmann2016}. We also observed \ng in \hstv (3 epochs), \hsti (3 epochs), and \hsth (6 epochs), to obtain well-calibrated Cepheid PLRs based on combinations of these filters~\citep{Riess2016}. All these observations were dithered to enable hot-pixel and/or cosmic-ray rejection. The  per-epoch exposure times were 1050, 1100, 1100, and 1106~s for \hstws, \hstvs, \hstis, and \hsths, respectively. A detailed observation log is presented in Table~\ref{tab_obs}.

\begin{deluxetable}{cccccc}
\tablecaption{Observation Log\label{tab_obs}}
\tablewidth{0pt}
\tablehead{
\colhead{Epoch} & \colhead{MJD} & \multicolumn{4}{c}{Dither$\times$Exposure Time (seconds)} \\ \cline{3-6}
&&\hstw & \hstv & \hsti & \hsth
}
\startdata
  1 &   57363.1 & 3$\times$350 & \nodata & \nodata & 2$\times$553 \\
  2 &   57370.7 & 3$\times$350 & 2$\times$550 & \nodata & \nodata \\
  3 &   57377.9 & 3$\times$350 & \nodata & 2$\times$550 & \nodata \\
  4 &   57384.7 & 3$\times$350 & \nodata & \nodata & 2$\times$553 \\
  5 &   57390.0 & 3$\times$350 & 2$\times$550 & \nodata & \nodata \\
  6 &   57392.8 & 3$\times$350 & \nodata & \nodata & 2$\times$553 \\
  7 &   57399.1 & 3$\times$350 & \nodata & 2$\times$550 & \nodata \\
  8 &   57405.9 & 3$\times$350 & \nodata & \nodata & 2$\times$553 \\
  9 &   57414.7 & 3$\times$350 & 2$\times$550 & \nodata & \nodata \\
 10 &   57423.6 & 3$\times$350 & \nodata & \nodata & 2$\times$553 \\
 11 &   57429.5 & 3$\times$350 & \nodata & 2$\times$550 & \nodata \\
 12 &   57436.9 & 3$\times$350 & \nodata & \nodata & 2$\times$553
\enddata
\end{deluxetable}\vspace{-0.5cm}

\subsection{Data Reduction}

We retrieved calibrated images (flat-fielded, dark-subtracted, charge-transfer efficiency corrected, etc.) from the Mikulski Archive for Space Telescopes (MAST), then registered and drizzled them using the {\tt AstroDrizzle v2.2.6} package. All the optical images (\hstws, \hstvs, and \hstis) were iteratively registered to the first \hstw image, while all the NIR images were registered to the first \hsth image due to the different optics and pixel scales of the IR and UVIS detectors. We drizzle-combined the aligned images of each band into deep master images, with cosmic rays rejected for the optical bands. We also produced a single drizzled image for each epoch, which serves as the basis for the time-series photometry. We set the drizzle pixel scales to the native  0.04\arcsec\ for the optical images, and a 1/3-finer-than-native (i.e., 0.08\arcsec/pixel) for the \hsth images.

\begin{figure}
\epsscale{1.2}
\plotone{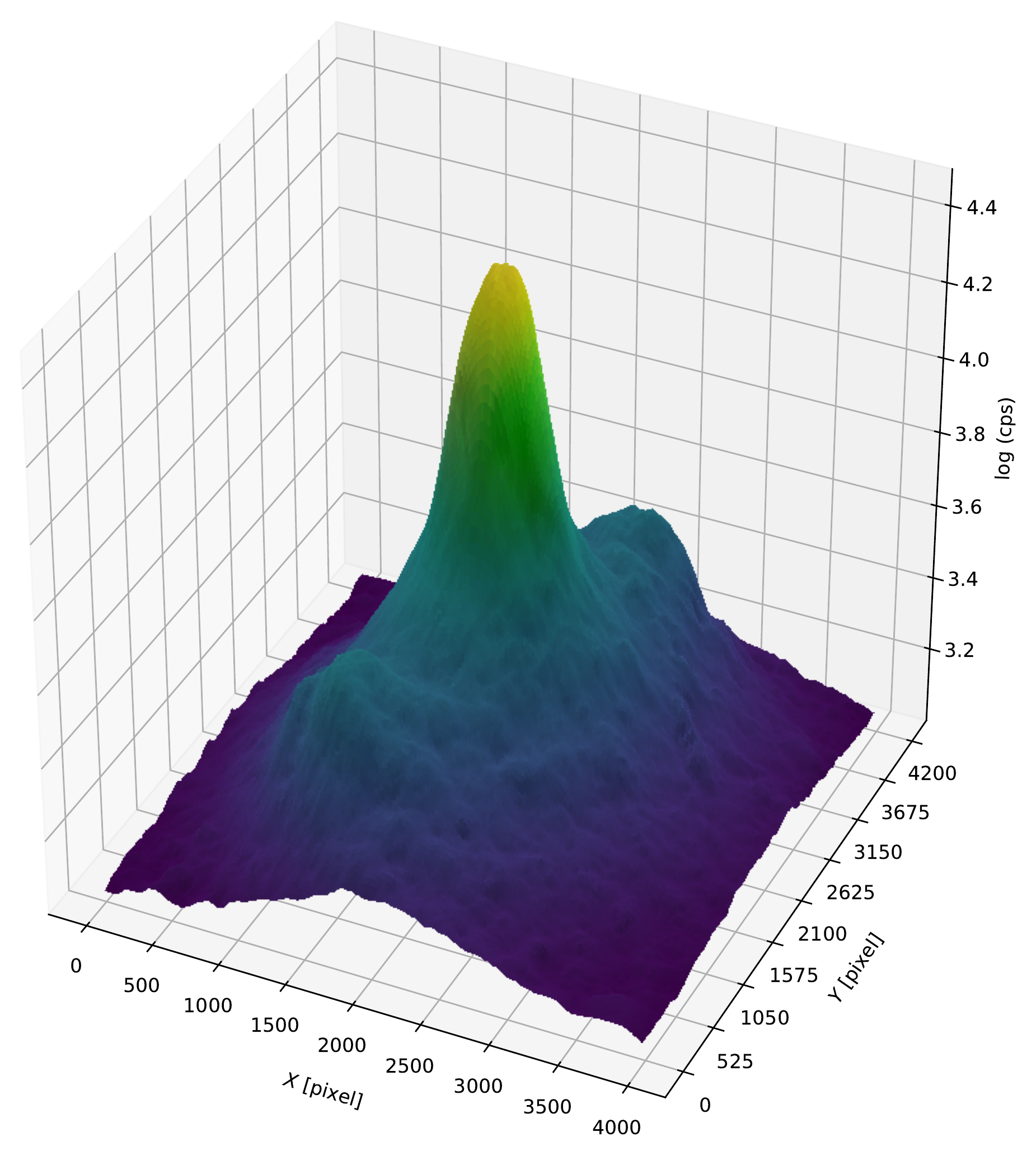}
\caption{The \hstw surface brightness model shows large gradient and complex variations of the background level across \ngs, in units of count rate per square arcsecond. Note the $z$-axis is plotted using a log scale.\label{fig_sbf}}
\end{figure}

We modeled the surface brightness gradients of all the drizzled images and subtracted these models prior to carrying out the photometry. \ng exhibits a high surface brightness in a large central region of the disk and a steep gradient in the outskirts, which hampers the ability of {\tt DAOPHOT} to detect sources and accurately fit PSFs. We masked all point sources found in each image at $>7\sigma$ significance, performed thin-plate spline fits to the background pixel counts across the image using the {\tt R} implementation of the {\tt Tps} function with default settings, and obtained a smooth surface brightness model. Fig.~\ref{fig_sbf} shows the results for \hstw as an example. These models were also used for local crowding estimation for Cepheids, as detailed below.

\subsection{Photometry}\label{sec_phot}

Given their superior depth and resolution, the \hstw data were used to derive the positions of point sources. This information was propagated to other bands for fixed-position photometry. This strategy not only allowed us to track stars in different bands with the same set of identifiers (IDs), but also provided more accurate flux measurements for other bands, especially in \hsth where the resolution is much worse. We summarize below the photometry procedures for each band.

\paragraph{\hstws} We started by creating a source list from a two-pass 3$\sigma$ source detection on the \hstw master image using {\tt DAOPHOT} and {\tt ALLSTAR}~\citep{Stetson1994}, then performed simultaneous PSF photometry on the 12 epochs of \hstw images using {\tt ALLFRAME}~\citep{Stetson1994}. In this step, we allowed both the stellar positions and flux to vary until a global PSF fit of all the sources converged. We excluded from the photometry results spurious sources associated with the stray light of foreground bright stars or background extended sources, as well as stars with exceptionally large (top 0.38\%) measurement errors for their magnitudes. We selected 57 bright, isolated, non-variable stars as secondary standards to calibrate the epoch-to-epoch magnitude zeropoint offsets. Their positions and calibrated Vega magnitudes are listed in Table~\ref{tab_sec}. We extracted the \hstw light curves using the {\tt TRIAL} program kindly provided by Peter Stetson.

\begin{deluxetable}{cccccc}
\tabletypesize{\scriptsize}
\tablecaption{Secondary Standards\label{tab_sec}}
\tablewidth{0pt}
\tablehead{
\colhead{ID} & \colhead{R.A.} & \colhead{Decl.} & \colhead{\hstw} & \colhead{\hstv} & \colhead{\hsti} \\ \cline{4-6}
& \multicolumn{2}{c}{[J2000]} & \multicolumn{3}{c}{[mag(mmag)]}
}
\startdata
    79502 &   182.61050 &    39.41246 &   26.237(17) &   27.647(23) &   27.210(22) \\
    73817 &   182.61060 &    39.41312 &   25.443(18) &   27.023(10) &   27.046(28) \\
    41386 &   182.61072 &    39.41676 &   24.514(25) &   26.120(14) &   25.980(19) \\
    76745 &   182.61148 &    39.41319 &   24.180(18) &   25.788(12) &   25.006(10) \\
    59463 &   182.61307 &    39.41580 &   25.214(12) &   27.507(20) &   25.217(14) \\
    37771 &   182.61457 &    39.41887 &   24.647(26) &   26.168(17) &   25.918(29) \\
   158737 &   182.61525 &    39.40575 &   23.987(21) &   25.580(13) &   25.294(17) \\
   149856 &   182.61543 &    39.40682 &   25.981(18) &   27.548(15) &   26.883(35) \\
     35238 &   182.61730 &    39.42037 &   25.395(16) &   26.940(19) &   26.669(27) \\
     118806 &   182.61890 &    39.41185 &   25.562(12) &   27.048(16) &   26.451(26) \\
     60500 &   182.61891 &    39.41829 &   24.209(23) &   25.890(11) &   25.572(16) \\
    202324 &   182.61935 &    39.40209 &   26.180(19) &   27.798(18) &   27.990(50) 
\enddata
\tablecomments{Coordinates are based on the WCS solution of the first \hstw image. Uncertainties are given in parentheses and are expressed in units of mmag. This table is available in its entirety in machine-readable form.}
\end{deluxetable}

\paragraph{\hstv \& \hstis} We performed \hstv and \hsti photometry using a modified version of {\tt ALLFRAME} which fixes the stellar positions during the PSF fitting. We ran {\tt ALLFRAME} on the \hstws, \hstvs, and \hsti master images, with positions fixed to those derived from the aforementioned \hstw time-series photometry. This method kept the stellar centroids from drifting during the PSF fit and provided high precision \hstv and \hsti flux measurements. We based the optical magnitude calibration on the photometric measurements from this step. Our analysis does not include time-series photometry for \hstv \& \hstis. Instead we phase-corrected the master-image photometry to mean magnitudes based on the \hstw light curves (see \S\ref{sec_cali} for details).

\paragraph{\hsths} While the \hstw stellar catalog contained a complete set of detectable sources in \hstv and \hstis, it did not include a large number of redder objects that appear only in the \hsth image. We simultaneously fit all the point sources in the \hsth image by combining the list of \hstw objects that were detected in \hsth with those that were found only in the NIR. We first transformed the positions of the \hstw catalog to the \hsth frame and performed a preliminary PSF fit. We then subtracted any detected stars from the \hsth master image and obtained a residual image. After that, we ran a two-pass 3$\sigma$ source detection on the residual image to find additional sources that only appeared in \hsths. Finally, we combined the lists to form a complete \hsth source input catalog and performed iterative PSF fitting using {\tt ALLSTAR} in fix-position mode to obtain the final \hsth photometry. As with \hstv and \hstis, we did not perform time-series photometry on the \hsth images.

\section{Cepheid Search}

We searched for Cepheids in \ng using a combination of template fitting and visual inspection.

\subsection{Variable Candidate Selection}

We selected candidate variables using the widely-used \citet{Stetson1996} $L$ index, which measures the significance of variability in a light curve taking measurement error into account. In our case, $L$ is reduced to
\begin{gather*}
    L = \frac{\sum\limits_i \mathrm{sgn}(\delta_i^2-1){|\delta_i^2-1|}^{1/2}}{0.798\cdot 12 \cdot \sqrt{n}} \cdot \frac{\sum\limits_i|\delta_i|}{\sqrt{\sum\limits_i\delta_i^2}}\\
    \delta_i = \sqrt{\frac{n}{n-1}}\frac{v_i-\overline{v}}{\sigma_i},
\end{gather*}
where {\tt sgn} is a function that returns $-1$ for negative values and 1 otherwise, $n$, $v_i$, $\sigma_i$, and $\overline{v}$ are the total number of measurements, the $i$th magnitude, the $i$th photometric uncertainty, and the weighted mean magnitude, respectively, for \hstws. Constant stars should have $L\sim 0$, while Cepheids should exhibit positive $L$~\citep{Stetson1996}.

\begin{figure}
\epsscale{1.2}
\plotone{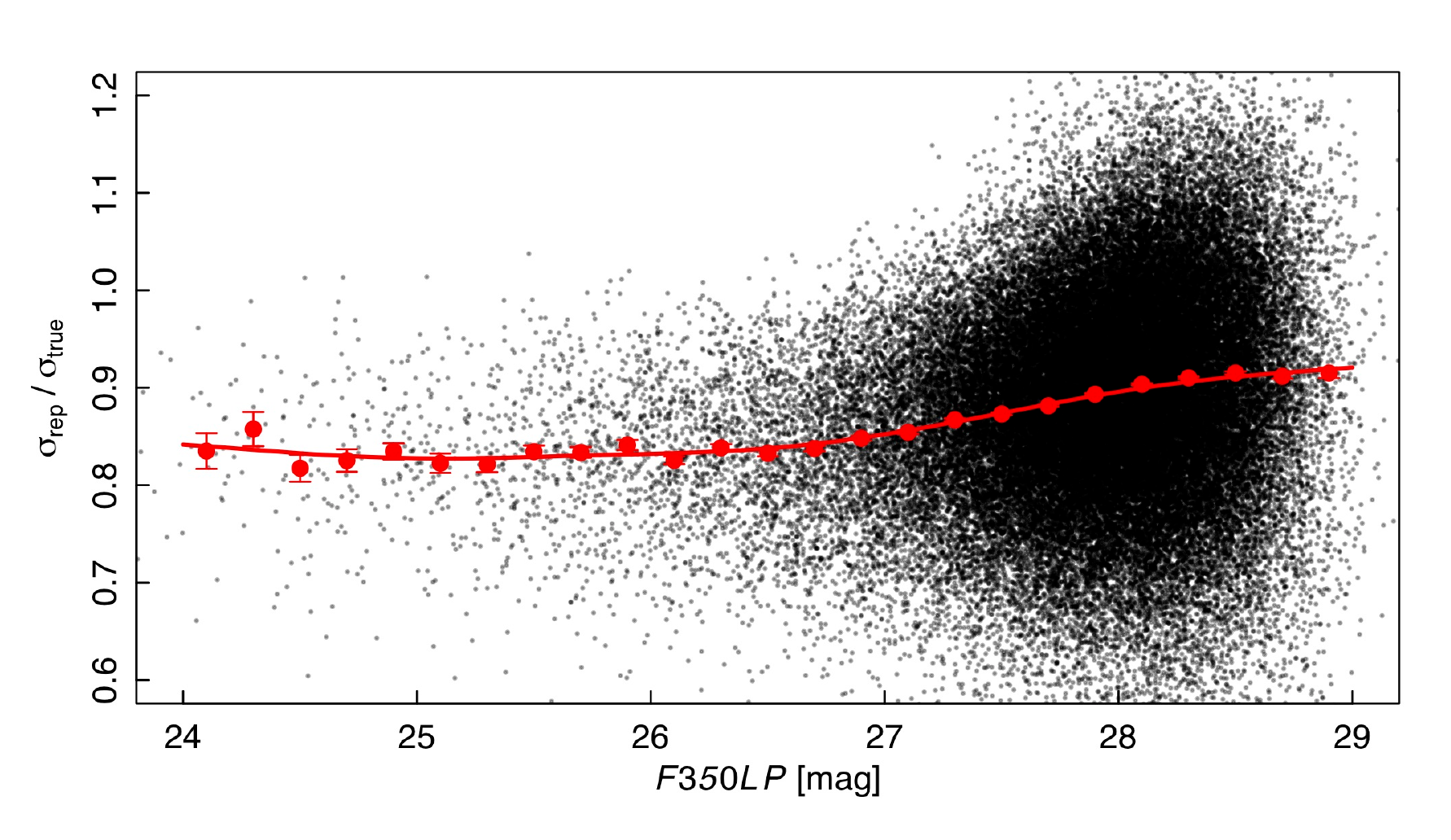}
\caption{Illustration of the magnitude-dependent errors in measurement errors for nonvariable stars ($-0.1<L<0.1$). The vertical axis represents the ratio of reported errors and true errors, obtained by rescaling the reduced $\chi^2$ to unity. The red points and curve indicate the median values in each magnitude bin and their spline fit, respectively.\label{fig_er}}
\end{figure}

\begin{figure}
\epsscale{1.2}
\plotone{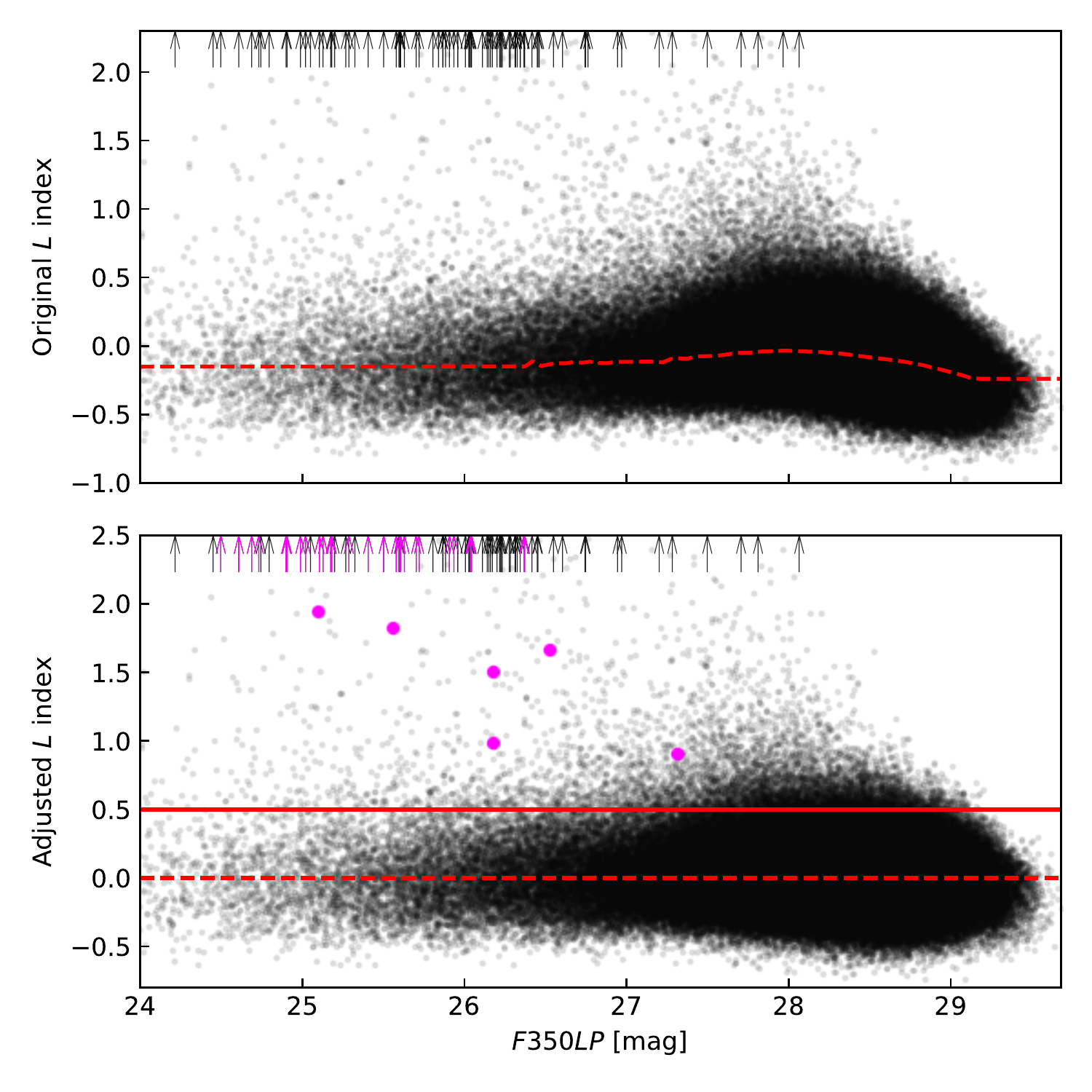}
\caption{The original (top) and adjusted (bottom) $L$ index against \hstw magnitudes. The red dashed lines indicate the median curves of the $L$--$m$ relations. The red solid line in the lower panel indicates our threshold for variable selection. Points with $L>2.5$ are indicated by arrows. Magenta symbols indicate the Cepheids described in \S\ref{sec_cep}. \label{fig_l2}}
\end{figure}

In practice, the photometric uncertainties reported by {\tt ALLFRAME} are known to exhibit nonlinear and magnitude-dependent biases (as illustrated in Fig.~\ref{fig_er}) which often lead to a nonuniform shape for the variability--magnitude ($L$--$m$) distribution~\citep{1998AJ....115.1016K}. Instead of rescaling measurement errors, we subtracted the median curve of the $L$--$m$ relation. The original and adjusted relations are shown in Fig.~\ref{fig_l2}. We conservatively chose $L>0.5$ and selected 6,956 variable candidates out of 403,738 objects. For reference, \citet{Stetson1996} adopted $L>0.9$ while \citet{Hoffmann2016} used $L>0.6-0.75$.

\begin{figure*}
\epsscale{1.2}
\plotone{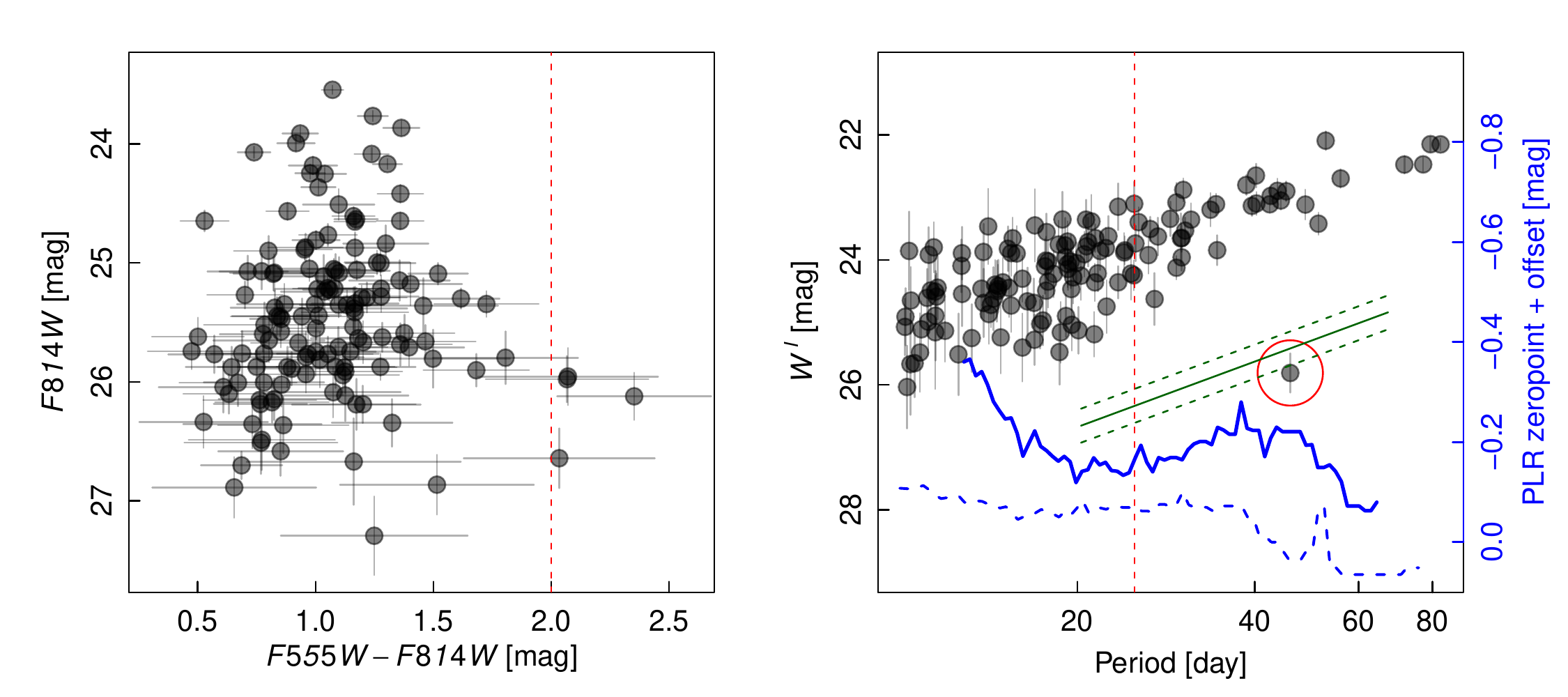}
\caption{Color (left) and period (right) cuts (shown as red dashed lines) imposed on the visually-identified Cepheid candidates. The variable identified by the red circle in the right panel, which is likely to be an RV Tauri variable, was excluded as well. Magnitudes and associated errors include crowding corrections in both plots. In the right panel, the solid and dashed blue curves, whose values should be read from the right-side blue axis, indicate PLR zeropoint variations for the boxcar test and minimum period cutoff test (see text), respectively. The solid and dashed green lines indicate the expected PLR and scatter of RV Tauri variables, based on the LMC sample of \citet{2008AcA....58..293S} transformed to the WFC3 bands following \citet{Riess2016}.\label{fig_prc}}
\end{figure*}

\subsection{Cepheid Template Fit}

\ \par

Thanks to their periodicity and predictable light curve shapes, Cepheids with sparse sampling can be accurately characterized with template-fitting methods~\citep[e.g.\ ][]{Stetson1996,2005MNRAS.363..749T}. We made use of the \citet{Yoachim2009} \V\ Cepheid templates, which are well applicable to the \hstw data given the similarity in effective wavelength between these two filters. 
\ \par
Similar to the practice of \citet{Shappee2011} and \citet{Hoffmann2016}, we modified the \citet{Yoachim2009} fitting method by letting the Cepheid amplitude be a free parameter. Their  templates were derived from eighth-order Fourier series, whose coefficients were firstly transformed into 4 leading parameters obtained from principal component analysis (PCA) and then modeled as polynomial functions of period. This yields fixed template amplitudes for a given period. Since Cepheids with the same period span a range of amplitudes~\citep{Stetson1996}, which are arguably sensitive to metallicity \citep[see][]{2012A&A...537A..81S,2013Ap&SS.344..381M}, we solved for the amplitude as a free parameter in our template-fitting procedure as neither well-calibrated metallicity--amplitude relation nor metallicity measurement for individual Cepheids is available for our case. The fitting model is expressed as

\begin{gather*}
  m  = \overline{m} + \frac{A}{\overline{A}(P)} \times \Big(\sum_{i=1}^8\alpha_i\sin (i\Phi) + \beta_i\cos (i\Phi) \Big)\\
  \Phi = 2\pi(\frac{t}{P}+\phi)\\
  \begin{pmatrix} \alpha \\ \beta \end{pmatrix} = \overline{\mathbf{V}} + \sum_{j=1}^4 a_j \cdot \mathbf{V}_j\\
  a_j = \mathrm{Polynomial}_j \big(\log P\big),
\end{gather*}
\vspace*{0.1cm}

\ \par

\noindent where $\overline{m}$, $A$, $P$, and $\phi$ are free parameters for mean magnitude, amplitude, period, and initial phase, respectively. $\overline{A}(P)$ is a period-dependent factor that scales the amplitudes of those original templates to unity. $\overline{\mathbf{V}}$ and $\mathbf{V}_j$ are the mean and $j$th PCA vector, respectively. Polynomial$_j$ represents the $j$th polynomial function as shown in Fig.~8 of \citet{Yoachim2009}. The allowed ranges of the free parameters were $0.01 < A < 2$ mag, $10 < P < 120$ days, and $0 < \phi < 1$.

We fit the templates to all the variable candidates and derived the best-fit parameters, despite the fact that most of them are not Cepheids. This is the reason we allow the fitting parameters to extend beyond the expected range of Cepheids in our template fitting procedure. We adopted the SLSQP method~\citep{kraft1988software} to search for the least square fit. To find the global minimum of the $\chi^2$, in each fit we used 1500 initial guesses for the nonlinear spaces $P$ and $\phi$, forming a meshgrid of 150 logarithmically spaced trial periods and 10 evenly spaced trial phases. We determined a template error of 0.05 mag using the OGLE Cepheid light curves~\citep{2008AcA....58..163S}, and added it to the covariance matrix in quadrature. We obtained the best-fit ``periods'' and ``amplitudes'' for all the candidate variables.

\ \par

We rejected objects exhibiting unrealistic amplitudes or periods that are longer than the observation baseline by 20\%, then visually inspected the remaining light curve fits to identify Cepheid candidates. We computed the \hstw amplitudes ($A$) of template fits, and rejected objects with $A>1.7$~mag or $A<0.4$~mag. Given the limited time span of the observations, we also restricted the visual selection sample to objects with ``periods'' shorter than 90 days. We visually inspected the light curves of 4,544 objects that survived these cuts and subjectively identified 136 Cepheid candidates that fit well to the saw-tooth shaped templates, with a philosophy that sample cleanliness is more important than completeness.

\subsection{Color \& Period Cuts}\label{sec_cep}

\begin{figure*}
\epsscale{1.15}
\plotone{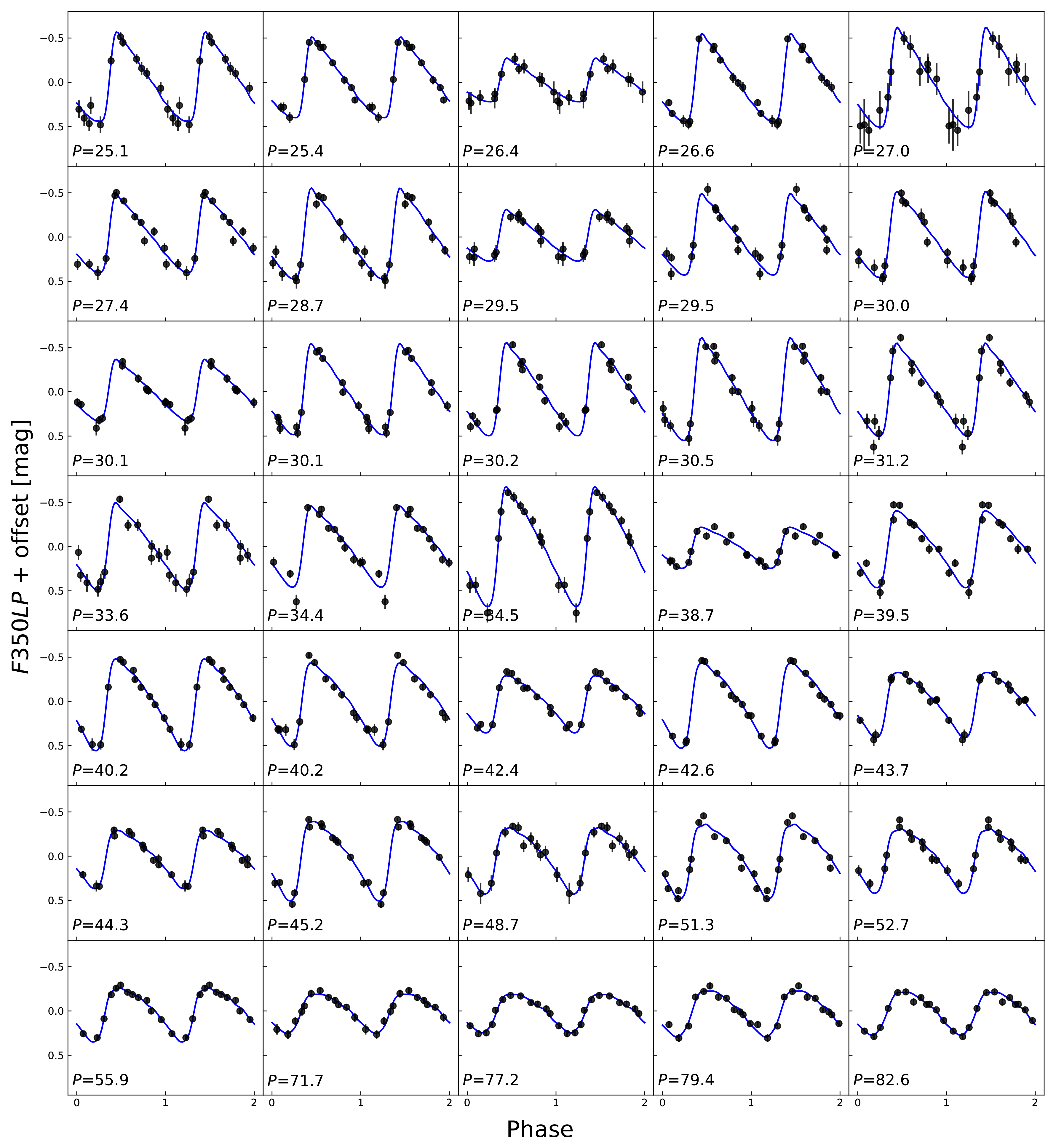}
\caption{\hstw light curves of the 35 optically identified Cepheids ordered by period. The blue curves indicate the best-fit models. We plot two cycles of pulsation for visualization purposes.\label{fig_lc}}
\end{figure*}

We inspected the $\hstv - \hsti$ colors and the optical Wesenheit PLR of the Cepheid candidates and applied additional cuts on color and period shown in Fig.~\ref{fig_prc}. We rejected 4 candidates with $\hstv - \hsti > 2$ mag, which are possibly due to extreme reddening or measurement errors. We found that the scatter of the optical Wesenheit PLR was significantly larger for $P\!~\!<25$\,d. Since the optical color is used in the NIR Wesenheit distance determination, we excluded 96 candidates with $P\!~\!<25$\,d. We further rejected one possible RV Tauri variable that is $\sim$8$\sigma$ off the classical Cepheid PLR, leaving 35 long period ($P\!>\!25$\,d) Cepheids that survived all cuts. We present their light curves in Fig.~\ref{fig_lc}. To investigate how the distance modulus is sensitive to the adopted period cutoff, we fit PLRs to subsamples of Cepheids with $\log (P)$ width of 0.2 dex running across the available period range like a ``boxcar.'' The PLR zeropoint variations of this running boxcar are shown in Fig.~\ref{fig_prc} with the blue solid curve. The steep drop in the $P\!~\!<20$\,d region is likely due to the incompleteness of low SNR Cepheids. The standard deviation of the zeropoint variation in the $P\!~\!>25$\,d region is 0.05 mag, consistent with the random error of the PLR zeropoint. We also tested the zeropoint variations as a function of minimum period cutoff, as shown in Fig.~\ref{fig_prc} with the blue dashed curve. Again, the standard deviation of their zeropoint variations is consistent with the random error of the PLR zeropoint. We find no detectable systematic distance error associated with the adopted $P\!~\!>25$\,d cutoff.

\section{Results}

As described in \S\ref{sec_phot}, our photometry strategy allowed us to track the \hsth Cepheid measurements by ID. We measured the \hsth magnitudes for 32 out of the 35 optically identified Cepheids, with one of them off the \hsth frame and two that were not detected in the \hsth frame. We computed their crowding bias in photometry for both \hsth and optical data using artificial star tests, and calibrated their magnitudes onto the Vega system. Finally, we obtained the optical and NIR Wesenheit PLRs of these Cepheids and derived the distance to \ngs.

\subsection{Crowding Correction}

We corrected for photometric bias including crowding through artificial star tests. ``Crowding'' refers to the measurement bias in crowded-field PSF photometry. This effect is location- and image-specific and highly dependent on the exact method used to carry out the photometry. It is usually characterized via artificial star tests~\citep[e.g., ][]{1988AJ.....96..909S,1996AJ....112.1928G,2006ApJS..166..534H}. We estimated the bias of our photometric procedure by injecting and measuring one artificial star at a time in the vicinity of each Cepheid. We repeated this process 100 times (200 times for \hsths) in order to obtain statistically solid  estimates. We also measured the dispersion of the recovered magnitudes of the artificial stars to obtain an estimate of the measurement uncertainty, which was added in quadrature to the Cepheid photometric errors. The magnitudes of artificial stars were based on the periods of their counterpart Cepheids and best-fit Cepheid PLRs, and as a result multiple realizations of the artificial star tests were performed to update the Cepheid PLRs with iteratively-determined corrections. We found mean corrections of 0.011, 0.006, 0.010, and 0.047~mag for \hstws, \hstvs, \hstis, and \hsths, respectively. The aim of the artificial star test is to measure the brightness of artificial stars in {\it exactly the same way} as for real stars, such that any bias in the real star photometry appears at the same level in the artificial star photometry. The specific size or scale of bias corrections are strongly dependent on the method and parameters used to measure the source photometry. Two of the most significant factors which impact the size of bias corrections include the minimum separation of nearest neighbor sources which may be de-blended from the source and the methods and statistics use to determine the local sky value. However, once corrected for the measured bias, accurate photometry should not be method-dependent. So we once again note that these crowding correction values are highly dependent on the adopted photometry strategy and thus cannot not be compared with crowding corrections obtained using different methods and/or applied to other data sets. The only meaningful comparison should be based on fully corrected and calibrated magnitudes for the same stars in the same images.

The term ``blending bias'' is sometimes interpreted differently from crowding bias \citep{2000AJ....120..810M,2001astro.ph..3440M}. While in practice crowding and blending can be highly degenerate or interchangeable \citep[e.g.,][]{2000PASP..112..177F,Riess2016}, the chance of physical associations between Cepheids and unresolved clusters or companion stars may lead to a bias if one assumes a uniform crowding level in the vicinity of a Cepheid where artificial stars are tested. \citet{2018ApJ...861...36A} cleared up some vague interpretations in earlier literature on this matter through quantitative studies of Cepheid companions and the physical association between Cepheids and open clusters, and found millimag-level biases on the distance modulus due to these effects. We thus excluded such an additional bias correction from our results.

\begin{figure}
\epsscale{1.2}
\plotone{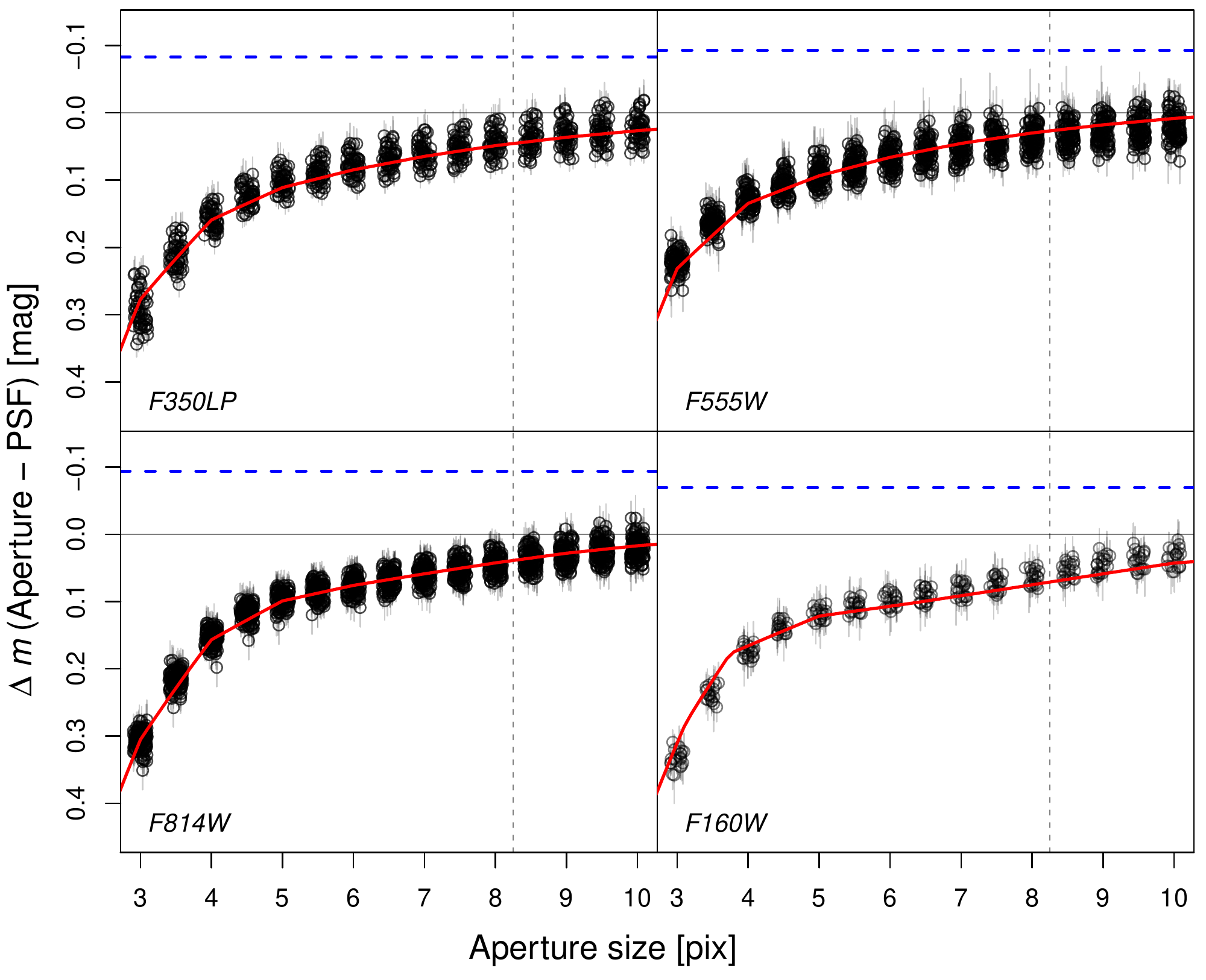}
\caption{Aperture corrections in \hstw, \hstv, \hsti and \hsth (clockwise from upper left). Circles indicate magnitude offsets between PSF and aperture photometry of those stars selected from SH0ES images (see text). Red lines are the growth curves obtained using {\tt PySynphot}. Blue lines indicates the correction to infinite aperture. Vertical dashed lines indicate the maximum radii used in the growth-curve fits.\label{fig_apc}}
\end{figure}

\subsection{Magnitude Calibration}\label{sec_cali}

We calibrated the Cepheid magnitudes onto the Vega system by adding the aforementioned crowding correction, as well as phase sampling correction, aperture correction, and Vega zeropoints, to be discussed below.

By convention, the mean magnitude of a Cepheid refers to its intensity mean over a full cycle. However, the directly measured magnitudes from the drizzle combined images are sparsely sampled and do not exactly cover one cycle. We computed the offset between the directly measured magnitude and full-cycle intensity mean using the best-fit \hstw template for each Cepheid. For the \hstv and \hsti band, we scaled the amplitude of the \hstw light curve by the relation in Table~2 of \citet{Hoffmann2016}. By simulating \hsth light curves (sinusoidal shape assumed) of a typical 0.3~mag amplitude, we found that the \hsth phase corrections are marginal due to the small amplitude, and thus we did not include them in the calibration. While based on the same idea, the exact mathematical treatment for our phase correction is slightly different from that of \citet{2013ApJ...764...84I} and \citet{2005PASP..117..823S}. In our case, we measured the Cepheid magnitudes based on stacked images (to take advantage of higher SNR than single-epoch images), which contain the integrated flux from multiple epochs. We computed the expected flux offset of a Cepheid between what is measured from the stacked image and the mean flux of a full template cycle, then added this offset back to the measured stacked flux to obtain the mean intensity. We determined 0.016 and 0.005 mag systematic errors associated with phase correction for $W^I$ and $W^H$, respectively.

\begin{deluxetable*}{cccccccccccrrr}
\tabletypesize{\scriptsize}
\tablecaption{Cepheid Properties\label{tab_ceph}}
\tablewidth{0pt}
\tablehead{
\colhead{ID} & \colhead{R.A.} & \colhead{Decl.} & \colhead{$P$} & \colhead{Ampl.} & \colhead{\hstw} & \colhead{\hstv} & \colhead{\hsti} & \colhead{\hsth} & SB$^c$ & F$^d$ & \colhead{$\Delta$\hstvs} & \colhead{$\Delta$\hsti} & \colhead{$\Delta$\hsth} \\ \cline{2-3} \cline{6-9}  \cline{12-14}
&\multicolumn{2}{c}{[J2000]$^a$}&[day]&[mag]&\multicolumn{4}{c}{[mag(mmag)]$^b$} & & & \multicolumn{3}{c}{[mmag(mmag)]$^e$}
}
\startdata
  365142 &   182.636887 &    39.389159 &    10.880 &     0.826 &    26.794( 83) &    26.986(170) &    26.171(172) &        \nodata &    210.6 & s &   -9( 95) &   22(142) &   \nodata \\
  367447 &   182.638440 &    39.389584 &    16.919 &     0.719 &    26.625( 72) &    26.732( 98) &    26.101(166) &        \nodata &    224.0 & s &  -13( 77) &   16(109) &   \nodata \\
  391593 &   182.645567 &    39.390032 &    10.225 &     0.663 &    26.860(110) &    26.879(129) &    26.024(185) &        \nodata &    248.7 & s &  -24(107) &   31(151) &   \nodata \\
  391936 &   182.649235 &    39.391621 &    21.427 &     1.017 &    25.830( 88) &    25.970(113) &    25.271(113) &        \nodata &    281.6 & s &  -28( 92) &   -1( 99) &   \nodata \\
  384416 &   182.647476 &    39.391687 &    12.753 &     0.534 &    26.609(106) &    26.751(119) &    25.793(214) &        \nodata &    273.5 & s &    8( 99) &   33(191) &   \nodata \\
  388289 &   182.649585 &    39.392186 &    15.783 &     0.660 &    26.440(138) &    26.582(149) &    25.415(171) &        \nodata &    294.3 & s &   -7(130) &   19(157) &   \nodata \\
  395256 &   182.652095 &    39.392541 &    25.412 &     0.914 &    25.535( 63) &    25.819( 81) &    24.766( 93) &    23.877(483) &    292.1 & h &   11( 68) &   42( 84) &  202(467) \\
  374104 &   182.646879 &    39.392583 &    33.649 &     0.987 &    26.446( 53) &    26.917(140) &    25.300( 85) &    23.734(324) &    280.4 & h &   -8( 38) &  -12( 57) &    8(316) \\
  336764 &   182.638068 &    39.392912 &    44.275 &     0.657 &    25.121( 39) &    25.378( 42) &    24.366( 60) &    23.588(223) &    252.0 & h &    4( 34) &   11( 38) &   20(204) \\
  380637 &   182.650438 &    39.393431 &    14.158 &     0.588 &    26.369(130) &    26.449(155) &    25.761(188) &        \nodata &    309.9 & s &  -15(146) &   25(163) &   \nodata
\enddata
\tablecomments{$a$: Coordinates are based on the WCS solution of the first \hstw image.
$b$: Fully calibrated Vega magnitudes, including crowding corrections and their uncertainties.
$c$: \hsths-band local surface brightness in the units of counts/s/sq arcsec.
$d$: Flag (including s, h, r, o, and c) indicating Cepheids rejected due to short periods (s; see \S3.3), used for $W^H$ PLR (h), rejected due to high SB (r), possible RV Tauri variable (o), or rejected due to red color (c).
$e$: Crowding corrections with uncertainties shown in parentheses.
This table is available in its entirety in machine-readable form.}
\end{deluxetable*}

\begin{figure*}
\epsscale{1.1}
\plotone{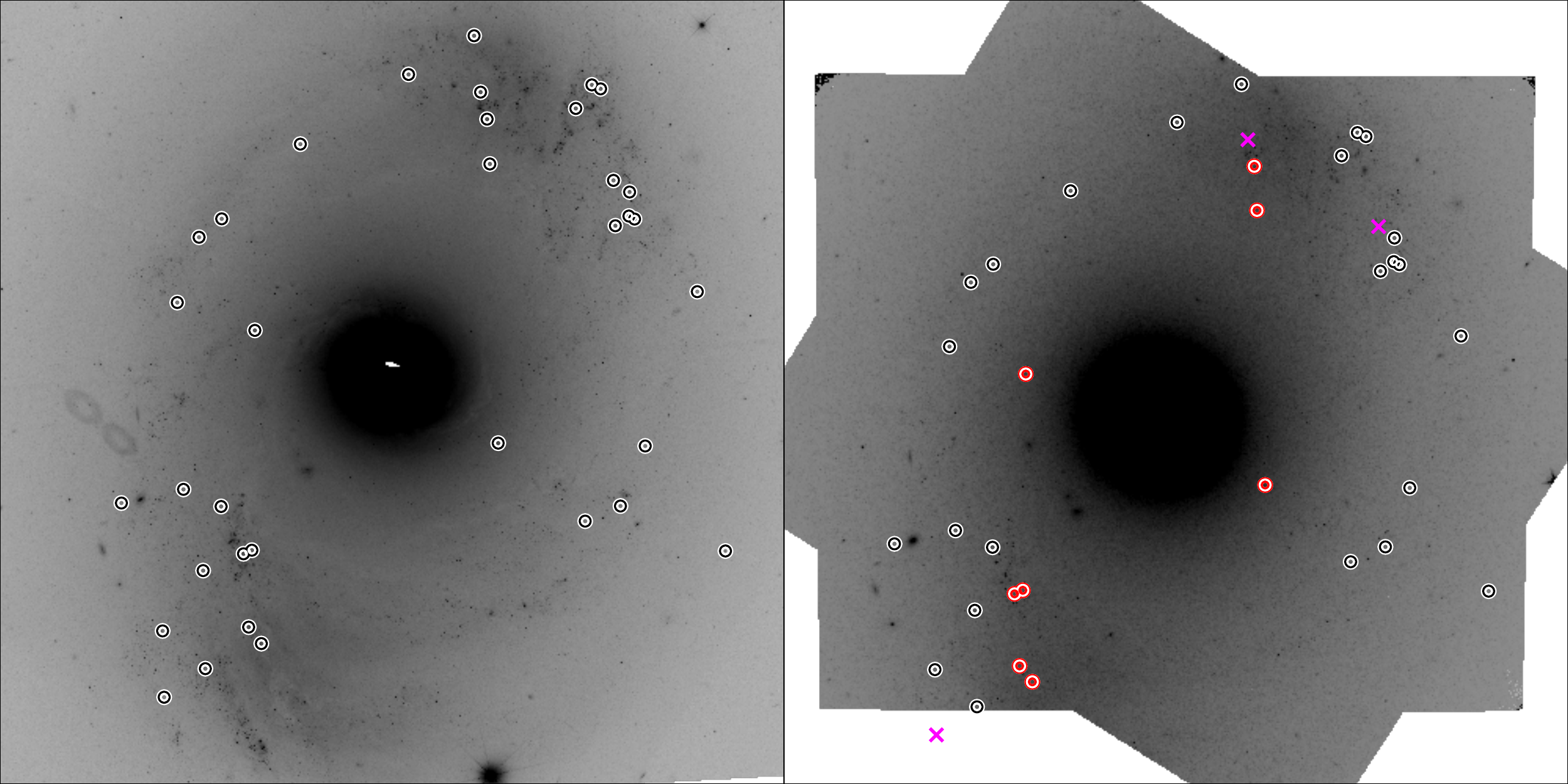}
\caption{Cepheid locations on the \hstw (left) and \hsth (right) master images. The red circles indicate Cepheids with \hsth local surface brightness beyond 300 counts/s/sq arcsec. Magenta crosses indicate Cepheids without valid \hsth photometry. \label{fig_loc}}
\end{figure*}

We derived the aperture correction and Vega zeropoints using the {\tt PySynphot v0.9.12} package~\citep{2013ascl.soft03023S}. Due to the lack of isolated bright stars in the \ng field, we derived PSF models and their aperture corrections based on \hst images taken with a similar configuration. We selected hundreds of suitable stars from the images acquired by the SH0ES team \citep{Riess2016}, who observed Cepheids in dozens of nearby galaxies. For \hsths, we made use of archival observations of the standard star P330E to construct the PSF model. We measured the aperture magnitudes of a series of aperture sizes using {\tt DAOPHOT}, and computed their offsets against the PSF magnitudes, as shown in Fig.~\ref{fig_apc}. These offsets were then fit to the growth curve obtained from {\tt PySynphot} by adjusting only the $Y$-intercept of those red curves in Fig.~\ref{fig_apc}. Since the offset between the magnitude measured in a certain aperture and an infinite aperture is fixed, and the offset between instrumental and Vega magnitudes is well established for these instruments and filters, we were able to convert the instrumental PSF magnitudes to infinite-aperture Vega magnitudes. We adopted a conservative 0.02 mag systematic uncertainty since we used the average of many calibrating stars in the aforementioned SH0ES fields to calibrate a different field, \ngs.

We list the properties and fully calibrated magnitudes of our Cepheid sample, including the rejected candidates, in Table~\ref{tab_ceph}. The locations of the Cepheids in the final sample are shown in Fig.~\ref{fig_loc}.

\begin{figure*}
\epsscale{1.2}
\plotone{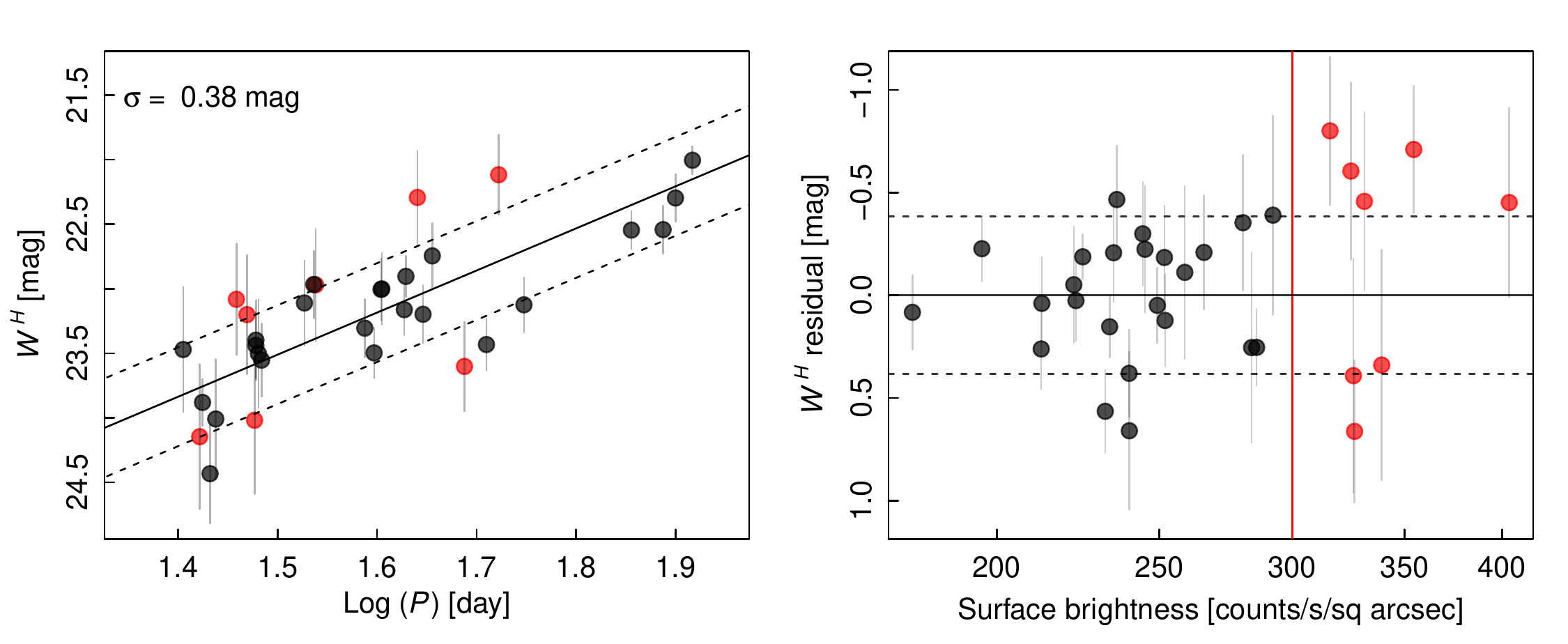}
\caption{$W^H$ PLR for all Cepheids with NIR measurements (left) and their residuals against local surface brightness (SB, right). The black solid and dashed lines indicate the PLR fit and $\pm1\sigma$ scatter, respectively. The red points indicate Cepheids rejected by the SB cut, indicated by the vertical red line.\label{fig_sb}}
\end{figure*}

\begin{figure*}
\epsscale{1.15}
\plotone{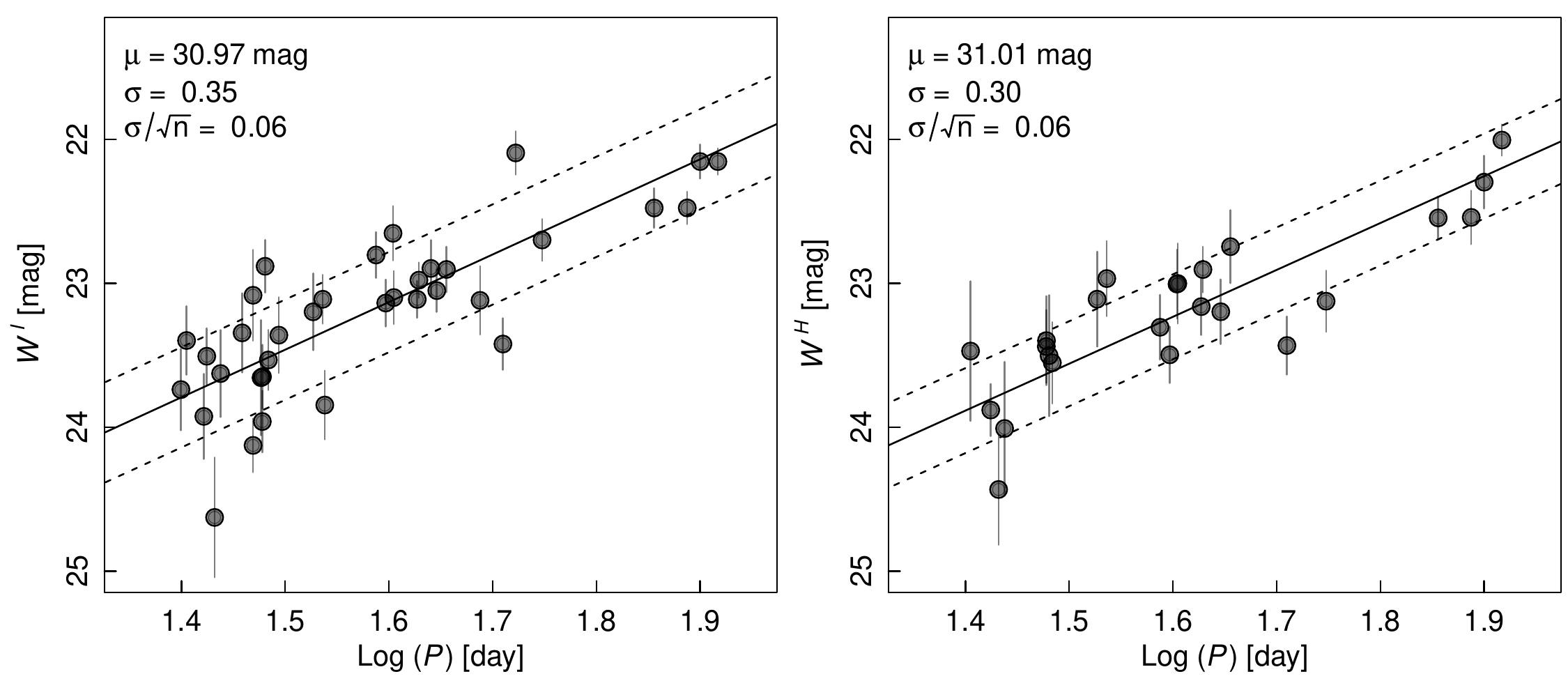}
\caption{$W^I$-band (left) and $W^H$-band (right) PLRs of the final Cepheid sample. The black solid and dashed lines indicate the PLR fit and $\pm1\sigma$ scatter, respectively. The distance moduli are based on the LMC PLRs of \citet{Riess2019} and the LMC distance determined by \citet{2019Natur.567..200P}. \label{fig_plr}}
\end{figure*}

\begin{deluxetable*}{cllccccc}
\tabletypesize{\small}
\tablecaption{Cepheid PLRs\label{tab_plr}}
\tablewidth{0pt}
\tablehead{
\colhead{Index} & \multicolumn{2}{c}{Expression} & \colhead{Slope} & \multicolumn{2}{c}{Intercept}  & \colhead{Scatter} & \colhead{$N$} \\ \cline{5-6}
&& & & LMC & \ng &
}
\startdata
$W^I$ & $\hsti - 1.3$   & $(\hstv-\hsti)$ & -3.31 & 15.933 & 28.426 & 0.348 & 35\\
$W^H$ & $\hsth - 0.386$ & $(\hstv-\hsti)$ & -3.26 & 15.915$^*$ & 28.449 & 0.295 & 24
\enddata
\tablecomments{$*$: Intercept was determined based on the data in Table~2 of \citet{Riess2019} which includes the CRNL term. It is greater than the corresponding value in Table~3 of \citet{Riess2019} by $\sim$ 0.03 mag because that table did not include the CRNL term despite its note to the contrary (Riess, private communication).}
\end{deluxetable*}

\subsection{Cepheid PLRs}

We adopted the Wesenheit indices and Cepheid PLR calibrations of \citet{Riess2019}, who used the same combination of filters. Since the crowding level varies significantly across our field, we inspected the PLR residuals against crowding using local surface brightness ($SB$) as an indicator, as shown in Fig.~\ref{fig_sb}. We observed an increased scatter at high $SB$ levels, and thus restricted our sample to objects with $SB(\hsths) < 300$~counts/s/sq arcsec for the $W^H$ PLR. This surface brightness cut reduced the scatter in the $W^H$ PLR by 23\%. A cut based on the crowding correction {\it uncertainty} ($\sigma_\mathrm{cor}$) gives a result consistent with the $SB$ cut to the level of PLR random error. We tested the PLR zeropoint variation for various $\sigma_\mathrm{cor}$ cuts from 0.15 mag to 0.4 mag, and obtained a standard deviation of 0.05 mag for the resultant distance moduli, which is consistent with the random error of the PLR. We concluded that the systematic error due to crowding correction, if it exists, is smaller than the random error, and excluded such a term in the final error budget. We present the final optical ($W^I$) and NIR ($W^H$) PLRs in Fig.~\ref{fig_plr} and in Table~\ref{tab_plr}. We note that the \hsth count-rate non-linearity correction is included in the LMC calibration of \citet{Riess2019}.

\subsection{Distance to \ngs}

We adopted the LMC as the single anchor for this work, as both its Cepheid PLRs and absolute distance have the highest precision to date~\citep{Riess2019, 2019Natur.567..200P}. We fixed the slopes of the PLRs to those derived by \citet{Riess2019} and computed the PLR intercept offsets between the LMC and \ngs. We derived relative distance moduli of $\Delta\mu=12.493\pm0.072$ and $\Delta\mu=12.534\pm0.074$\,mag using the $W^I$ and $W^H$ PLRs, respectively, and a weighted mean of $\Delta\mu=12.51\pm0.05$\,mag. Adding the geometrically determined LMC distance modulus of $\mu=18.477\pm 0.026$\,mag~\citep{2019Natur.567..200P}, we obtained a distance modulus for \ng of $\mu=30.99\pm0.06$ mag ($D=15.8\pm0.4$ Mpc).

The leading error sources in our distance determination are: the random error due to PLR scatter, the sensitivity of the zeropoint on the adopted PLR slope, the uncertainty in the absolute distance to the LMC, the aperture corrections (see \S\ref{sec_cali}) and the metallicity dependence of Cepheid PLRs. We present the full error budget in Table~\ref{tab_err}.

To estimate the distance error due to metallicity, we multiplied the scatter of [O/H] measurements in 21 galaxies presented by \citet{Riess2016} by the metallicity term coefficients ($-0.20$ and $-0.13$ per dex of [O/H] for $W^I$ and $W^H$, respectively) in Equation~2 of \citet{Riess2016}. Due to the lack of consistent [O/H] measurements for this system, we did not apply a metallicity correction for \ngs.

\begin{figure*}
\epsscale{1.2}
\plotone{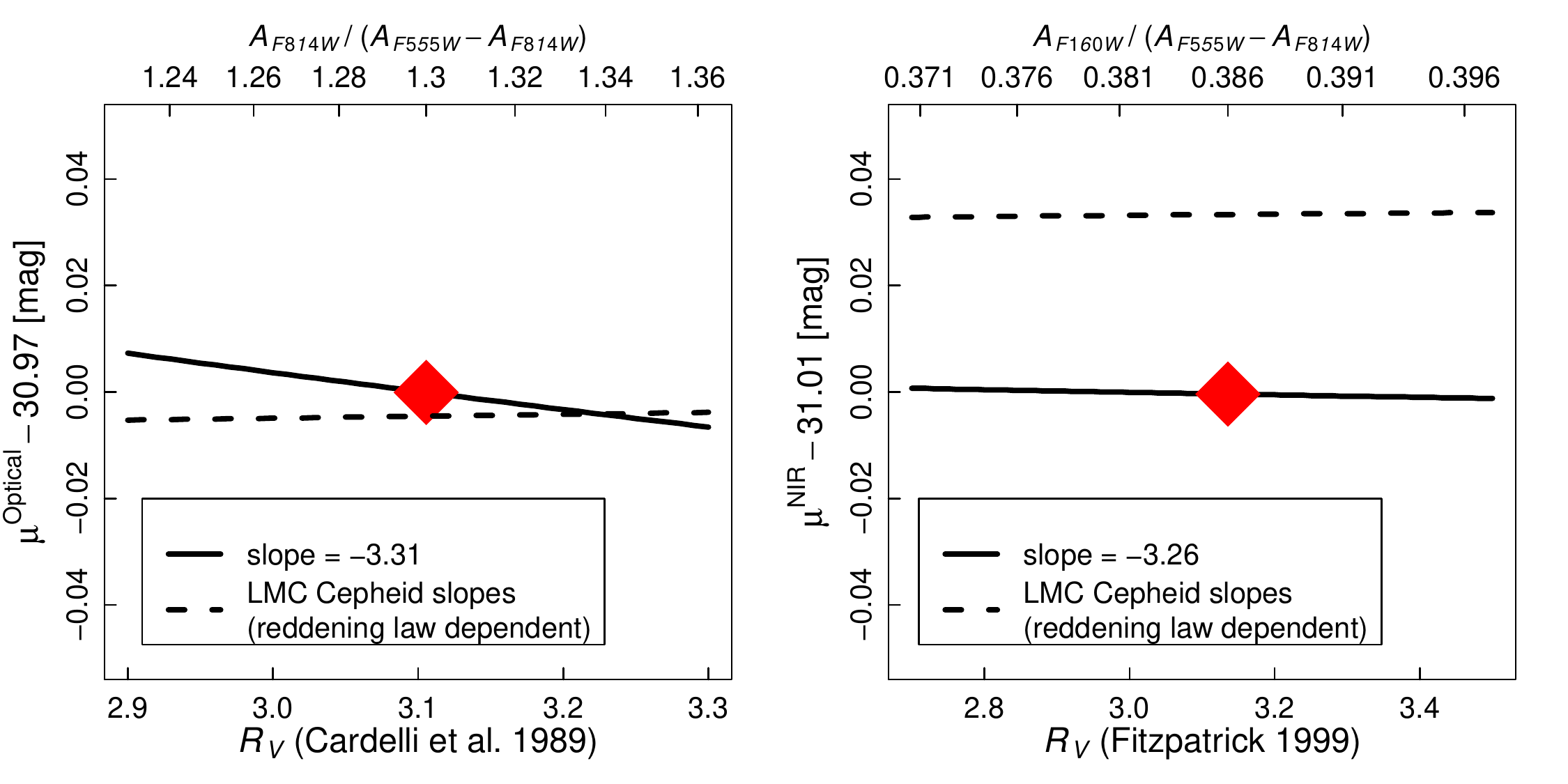}
\caption{Distance dependence on reddening laws (solid lines) and PLR slopes (dashed lines) for $W^I$ (left) and $W^H$ (right). The reddening laws are expressed in both $R_V$ values (bottom axis) and Wesenheit color terms (top axis). The adopted reddening laws are indicated by red diamonds. The 0.033 mag offset of the dashed line in the right panel is due to the steeper PLR in the LMC than the average PLR slope of a broader sample of galaxies, which is given by \citet{Riess2019}.\label{fig_red}}
\end{figure*}

We investigated the systematic uncertainties due to reddening law by testing a range of $R_V$ values. For the optical Wesenheit index $W^I$, we computed the color term coefficient $A_\hsti / E(\hstv - \hsti)$ for $R_V$ ranges from 2.9 to 3.3 by convolving the \citet{1989ApJ...345..245C} extinction curve with the WFC3 filter transmission functions, then applied them to both the LMC and \ng Cepheid samples with PLR slope fixed to the \citet{Riess2019} choice. Because only the differential effect between the two galaxies contributes to the \ng distance calibration, the systematic uncertainty due to the choice of reddening law is secondary, as shown in the left panel of Fig.~\ref{fig_red}. We adopted a conservative uncertainty of 0.01 mag for the $W^I$-based distance. We performed the same test for the NIR Wesenheit index $W^H$ with reddening laws of $2.7<R_V<3.5$ using \citet{1999PASP..111...63F}, the same formulation adopted by \citet{Riess2016,Riess2019}. As expected, we found an even weaker distance dependence on the reddening law for $W^H$, as shown in the right panel of Fig.~\ref{fig_red}.

We studied the \ng distance dependence on the PLR slopes by redetermining the PLR slopes for various reddening laws using only the LMC sample studied by \citet{Riess2019}. We excluded the same outliers (OGL0992 and OGL0847) as \citet{Riess2019} in our calculations, though we note including them did not change the results. We found steeper $W^H$-band slopes than \citet{Riess2019}, who constrained the $W^H$ PLR slope with a much broader Cepheid sample. The distance variations of using these LMC sample constrained and reddening law dependent slopes are shown with dashed lines in Fig.~\ref{fig_red}. We adopted the 0.033 mag offset due to steeper $W^H$-band slopes as a systematic error on the $W^H$-based distance modulus.

\begin{deluxetable}{lll}
\tabletypesize{\normalsize}
\tablecaption{Error Budget\label{tab_err}}
\tablewidth{0pt}
\tablehead{
\colhead{Source} & \colhead{$W^I$} & \colhead{$W^H$} \\ 
& (mag) & (mag)
}
\startdata
PLR scatter & 0.06 & 0.06 \\
PLR slope & \nodata & 0.033 \\
LMC distance & 0.026 & 0.026 \\
Magnitude calibration & 0.02 & 0.02 \\
Metallicity & 0.03 & 0.02 \\
Phase correction & 0.016 & 0.005 \\
Reddening law & 0.01 & \nodata \\
\hline
Total & 0.077 & 0.079
\enddata
\end{deluxetable}

\section{Discussion}

\subsection{Black Hole Masses}

One of the primary reasons for undertaking this investigation has been to facilitate comparisons among various methods of determining the masses of the central black holes in these galaxies. NGC\,4151 and NGC\,3227 are among a handful of galaxies for which the black hole radius of influence $r_{\rm BH} = GM_{\rm BH}/\sigma_*^2$,  where $\sigma_*$ is the stellar velocity dispersion, is resolved so masses based on stellar or gas dynamical modeling have high credibility. An accurate distance is required since masses based on stellar or gas dynamics are distance dependent. Masses based on reverberation mapping are, by contrast, distance-independent. Reverberation-based masses are computed from
\begin{equation}
M_{\rm BH} = f \left( \frac{\Delta V^2 c \tau}{G} \right),
\end{equation}
where $\tau$ is the reverberation time lag between continuum and emission-line flux variations, $\Delta V$ is the emission line width, and $f$ is a dimensionless scaling factor of order unity that depends on the inclination, geometry, and kinematics of the broad emission-line region. The scale factor $f$ can be determined by detailed modeling of reverberation data \citep[e.g.,][]{Pancoast2014a,Grier2017,Horne2020} and seems to be most sensitive to inclination \citep{Grier2017}. In the absence of detailed modeling, it is customary to use an average scaling factor \citep{2004ApJ...615..645O} that is determined by assuming that both quiescent and active galaxies follow the same relationship between black hole mass $M_{\rm BH}$ and bulge velocity dispersion $\sigma_*$ \citep{Ferrarese2000,Gebhardt2000,Gultekin2009}. The most recent calibration gives $\langle \log f \rangle = 0.683$ \citep{Batiste2017} with an error on the mean of $0.030$\,dex and a standard deviation of $0.150$~dex. Since detailed reverberation models of NGC\,3227 and NGC\,4151 are not yet available, we use this value here, and note that this introduces a systematic uncertainty of $\sim0.150$\,dex.

In Table \ref{table:mass_comp}, we show various mass determinations for both NGC\,3227 and NGC\,4151  along with assumed distances for each investigation. Our preferred distance to NGC\,3227 is based on a surface brightness fluctuation measurement of its interacting elliptical companion, NGC\,3226, as determined by \cite{Tonry2001} with a correction by \cite{Blakeslee2010}. Table \ref{table:mass_comp} also shows mass measurements for the central black hole in both galaxies adjusted to our preferred distances of 23.7\,Mpc for NGC\,3227 and 15.8\,Mpc for NGC\,4151.

\begin{deluxetable}{lcccc}
\tabletypesize{\scriptsize}
\tablecaption{Black Hole Mass Comparison\label{table:mass_comp}}
\tablewidth{0pt}
\tablehead{
\colhead{Source} & 
\colhead{Ref.} & 
\colhead{Method} &
\colhead{$D$ (Mpc)} &
\colhead{$\log M/M_\odot$} \\
\colhead{(1)} &
\colhead{(2)} &
\colhead{(3)} &
\colhead{(4)} &
\colhead{(5)} 
}
\startdata
NGC\,3227 & 1 & stars & $17.0$ & $7.18^{+0.12}_{-0.33}$ \\
NGC\,3227 & 2 & gas   & $15.6$ & $7.30^{+0.18}_{-0.10}$ \\
NGC\,4151 & 3 & stars & $13.9$ & $7.58^{+0.12}_{-0.16}$ \\
NGC\,4151 & 2 & gas   & $13.2$ & $7.48^{+0.10}_{-0.58}$ \\
\hline 
NGC\,3227 & 1,4,5 & stars & $23.7$   & $7.32^{+0.12}_{-0.33}$ \\
NGC\,3227 & 2,4,5 & gas   & $23.7$   & $7.48^{+0.18}_{-0.10}$ \\
NGC\,3227 & 6     & RM    & \nodata & $6.65\pm 0.20 (\pm 0.15)$ \\
\hline
NGC\,4151 & 3,7   & stars & $15.8$   & $7.63^{+0.12}_{-0.16}$ \\
NGC\,4151 & 2,7   & gas   & $15.8$   & $7.56^{+0.10}_{-0.58}$ \\
NGC\,4151 & 7,8   & gas   & $15.8$   & $7.67^{+0.09}_{-0.11}$ \\
NGC\,4151 & 6     & RM    & \nodata & $7.61 \pm 0.05 (\pm 0.15) $
\enddata
\tablecomments{Columns are:
1: Source;
2: References;
3: ``stars'' refers to stellar dynamical modeling, ``gas'' refers to
gas dynamical modeling, ``RM'' refers to reverberation mapping;
4: Distance assumed;
5: Black hole mass.\\
The top part of this table contains the original mass measurements assuming the distances listed, while the two lower parts show the adjusted masses.}
\tablerefs{
1: \citet{Davies2006};
2: \citet{Hicks2008};
3: \citet{Onken2014};
4: \citet{Tonry2001} (distance);
5: \citet{Blakeslee2010} (distance correction);
6: \citet{DeRosa2018};
7: this work (distance);
8: Hicks et al., in preparation (mass).}
\end{deluxetable}

The mass measurements for the SMBH in NGC\,4151 include a recent determination based on gas dynamics using integral field spectroscopy (Hicks et al., in preparation) and the various techniques applied to this system are in excellent agreement. On the other hand, even with the currently large systematic uncertainty in the reverberation-based masses, the corresponding estimate for NGC\,3227 is much lower than the values based on stellar or gas dynamics. However, these differences can be understood as an inclination effect; indeed, unification models \citep{Antonucci93} predict that Type 1 AGNs, like NGC\,3227 and NGC\,4151, are preferentially at lower rather than higher inclination. If $f \propto 1/\sin^2 i$, which is appropriate for a thin disk, this would imply $i \approx 11^\circ$, which is consistent with the $15^\circ$ inclination of the narrow-line region \citep[][though note in their paper the inclination given is the complement of the standard definition]{Fischer2013}, but not with the less-reliable inclination of $33^\circ \pm 2^\circ$ inferred from the Fe K$\alpha$ profile \citep{Patrick2012}. However, the inclination of the narrow-line region in NGC\,4151 is only $21^\circ$ \citep{Winge1999}, so a similar inclination correction introduces some tension in the mass comparison. The conclusion we draw from this is that velocity-resolved reverberation mapping is required for a meaningful mass comparison in individual sources because the unknown geometry and inclination of the BLR makes reverberation-based mass measurement using only lag and line width too uncertain.

\subsection{Other Implications}

We noted in the Introduction that a wide range of distances have been reported for NGC\,4151. The distance of $19.0^{+2.4}_{-2.6}$\,Mpc derived by \cite{Hoenig2014} by combining near-IR interferometry with dust reverberation is $\sim20$\% larger than our Cepheid-based distance, but statistically consistent. On the other hand, the dust-reverberation distance of $29.2 \pm 0.4$\,Mpc \citep{Yoshii2014} is nearly a factor of two larger than our measurement, probably because at least one assumption in the theoretical prediction of the dust torus radius is incorrect; indeed, of the 17 objects studied, only two of 16 yielded distances consistent within the errors with the standard luminosity distance (we will discuss a Cepheid-based distance for the remaining source, NGC\,4051, in a subsequent paper).

The improved distance allowed us to improve both mass and energetics in one of the nearest and best targets to study the physics of active nuclei on small scales. For example, \cite{Storchi-Bergmann2009} derived  masses of ionized and hot molecular gas. These masses scale as $D^2$, and our revised distance thus increases these masses by 41\%. Similarly, the mass of ionized gas in the outflow that was calculated as $3.1 \times 10^6$\,\msun\ becomes $4.4 \times 10^6$\,\msun. \cite{Storchi-Bergmann2010} derived the mass outflow rate and outflow power. These both scale with $D$, because the rates are calculated at a certain distance from the nuclear source; thus these quantities are proportional to $D^2/D$, and therefore should be increased by 19\%.

\ \par

\vspace*{-1cm}

\section{Summary}

We obtained multiband time-series observations of the archetypical Seyfert 1 galaxy \ng with the {\it Hubble Space Telescope} and identified over 130 Cepheid candidates in this system. We used a subset of 35 long-period ($P\!>\!25$d) Cepheids to derive a distance of $15.8\pm0.4$~Mpc using the Near-Infrared Cepheid Period--Luminosity Relation. 

The precise Cepheid-based distance enabled us to update previous dynamical mass determinations for the SMBH in this AGN, which yield consistent values of $\log M/\msun\sim 7.6$~dex. These, in turn, are in good agreement with mass estimates based on reverberation mapping.

\acknowledgments

We thank K.Z.~Stanek for helpful advice early in this project. We thank the anonymous referee and Albert Bosma for providing valuable comments and suggestions. MMF and BMP thank Tod Lauer for an enlightening conversation. Support for {\em HST} program GO-13765 was provided by NASA through a grant from the Space Telescope Science Institute, which is operated by the Association of Universities for Research in Astronomy, Inc., under NASA contract NAS5-26555. MV gratefully acknowledges support from the Independent Research Fund Denmark via grant number DFF 8021-00130. 

\bibliographystyle{aasjournal}
\bibliography{ref}

\end{document}